# Small Spans in Scaled Dimension

John M. Hitchcock
Department of Computer Science
Iowa State University
jhitchco@cs.iastate.edu

## Abstract

Juedes and Lutz (1995) proved a *small span theorem* for polynomial-time many-one reductions in exponential time. This result says that for language $A$ decidable in exponential time, either the class of languages reducible to $A$ (the lower span) or the class of problems to which $A$ can be reduced (the upper span) is small in the sense of resource-bounded measure and, in particular, that the degree of $A$ is small. Small span theorems have been proven for increasingly stronger polynomial-time reductions, and a small span theorem for polynomial-time Turing reductions would imply BPP $\neq$ EXP. In contrast to the progress in resource-bounded measure, Ambos-Spies, Merkle, Reimann, and Stephan (2001) showed that there is no small span theorem for the resource-bounded dimension of Lutz (2000), even for polynomial-time many-one reductions.

Resource-bounded scaled dimension, recently introduced by Hitchcock, Lutz, and Mayordomo (2003), provides rescalings of resource-bounded dimension. We use scaled dimension to further understand the contrast between measure and dimension regarding polynomial-time spans and degrees. We strengthen prior results by showing that the small span theorem holds for polynomial-time many-one reductions in the $-3^{\text{rd}}$-order scaled dimension, but fails to hold in the $-2^{\text{nd}}$-order scaled dimension. Our results also hold in exponential space.

As an application, we show that determining the $-2^{\text{nd}}$- or $-1^{\text{st}}$-order scaled dimension in ESPACE of the many-one complete languages for E would yield a proof of P $=$ BPP or P $\neq$ PSPACE. On the other hand, it is shown unconditionally that the complete languages for E have $-3^{\text{rd}}$-order scaled dimension 0 in ESPACE and $-2^{\text{nd}}$- and $-1^{\text{st}}$-order scaled dimension 1 in E.

## 1 Introduction

Resource-bounded measure [14] defines the relative size of classes of decision problems and has been used very successively to study polynomial-time reductions within exponential-time complexity classes. Measure-theoretic arguments were the first to show that for all $\alpha < 1$, every $\leq^{\text{P}}_{n^\alpha\text{-tt}}$-hard language for exponential time is exponentially dense [16]. The first plausible hypothesis on NP to separate the $\leq^{\text{P}}_{\text{m}}$ and $\leq^{\text{P}}_{\text{T}}$ reducibilities within NP came from resource-bounded measure [17].

The *degrees* and *spans* of languages under polynomial-time reductions have also been studied by several researchers using resource-bounded measure. For a reducibility $\leq^{\text{P}}_r$ and any $A \subseteq \{0,1\}^*$, the $\leq^{\text{P}}_r$-*lower span* of $A$ is the class $\text{P}_{\text{m}}(A)$ of all languages that are $\leq^{\text{P}}_r$-reducible to $A$, the $\leq^{\text{P}}_r$-*upper span of $A$* is the class $\text{P}_{\text{m}}^{-1}(A)$ of all languages to which $A$ is $\leq^{\text{P}}_r$-reducible, and the $\leq^{\text{P}}_r$-*degree of $A$* is the class $\text{deg}^{\text{p}}_{\text{m}}(A) = \text{P}_{\text{m}}(A) \cap \text{P}_{\text{m}}^{-1}(A)$. Juedes and Lutz [10] proved the following *small span theorem* for $\leq^{\text{P}}_{\text{m}}$ reductions in both E and in EXP. Here $\mu(\mathcal{C} \mid \mathcal{D})$ denotes the *measure of $\mathcal{C}$ within $\mathcal{D}$*, where $\mathcal{D}$ is a suitable complexity class. If $\mu(\mathcal{C} \mid \mathcal{D}) = 0$, then $\mathcal{C}$ is a negligible subset of $\mathcal{D}$.

**Theorem 1.1.** (Juedes and Lutz [10]) *Let $\mathcal{D} \in \{\mathrm{E}, \mathrm{EXP}\}$. For every $A \in \mathcal{D}$,*

$$\mu(\mathrm{P_m}(A) \mid \mathcal{D}) = 0$$

*or*

$$\mu(\mathrm{P_m^{-1}}(A) \mid \mathcal{D}) = 0.$$

*In particular, $\mu(\mathrm{deg_m^P}(A) \mid \mathcal{D}) = 0$.*

That is, at least one of the upper or lower spans of $A$ is small within $\mathcal{D}$. Using a result of Bennett and Gill [4], Juedes and Lutz [10] noted that strengthening Theorem 1.1 from $\leq_\mathrm{m}^\mathrm{P}$ reductions to $\leq_\mathrm{T}^\mathrm{P}$ reductions would achieve the separation $\mathrm{BPP} \neq \mathrm{EXP}$. However, small span theorems for reductions of progressively increasing strength between $\leq_\mathrm{m}^\mathrm{P}$ and $\leq_\mathrm{T}^\mathrm{P}$ have been obtained by Linder [11], Ambos-Spies, Neis, and Terwijn [3], and Buhrman and van Melkebeek [6].

Resource-bounded dimension was introduced by Lutz [12] as an effectivization of Hausdorff dimension [7] to investigate the fractal structure of complexity classes. Just like resource-bounded measure, resource-bounded dimension is defined within suitable complexity classes $\mathcal{D}$. For any complexity class $\mathcal{C}$, the *dimension of $\mathcal{C}$ within $\mathcal{D}$* is a real number in $[0, 1]$ and is denoted by $\dim(\mathcal{C} \mid \mathcal{D})$. If $\dim(\mathcal{C} \mid \mathcal{D}) < 1$, then $\mu(\mathcal{C} \mid \mathcal{D}) = 0$, but the converse may fail. This means that resource-bounded dimension is capable of quantitatively distinguishing among the measure 0 sets. With regard to the measure 0 sets in Theorem 1.1, Ambos-Spies, Merkle, Reimann, and Stephan [2] proved the following.

**Theorem 1.2.** (Ambos-Spies, Merkle, Reimann, and Stephan [2]) *For every $A \in \mathrm{E}$,*

$$\dim(\mathrm{deg_m^P}(A) \mid \mathrm{E}) = \dim(\mathrm{P_m}(A) \mid \mathrm{E}).$$

In particular, as $\dim(\mathrm{E} \mid \mathrm{E}) = 1$, the $\leq_\mathrm{m}^\mathrm{P}$-complete degree for E has dimension 1 within E. This implies that replacing "$\mu$" by "dim" in Theorem 1.1 makes the statement for E no longer true. In other words, there is no analogue of the small span theorem for dimension in E. Dimension in E cannot distinguish between lower spans and degrees.

To overcome limitations of resource-bounded dimension for investigating complexity classes within ESPACE, Hitchcock, Lutz, and Mayordomo [9] introduced for each integer $i \in \mathbb{Z}$ an $i^{\mathrm{th}}$-*order scaled dimension* $\dim^{(i)}(\cdot \mid \mathcal{D})$. For any class $\mathcal{C}$ and $i \in \mathbb{Z}$, $\dim^{(i)}(\mathcal{C} \mid \mathcal{D}) \in [0, 1]$, and if it is less than 1, then $\mu(\mathcal{C} \mid \mathcal{D}) = 0$. The quantity $\dim^{(i)}(\mathcal{C} \mid \mathcal{D})$ is nondecreasing in $i$, and there is at most one $i \in \mathbb{Z}$ for which $0 < \dim^{(i)}(\mathcal{C} \mid \mathcal{D}) < 1$. The $0^{\mathrm{th}}$-order dimension, $\dim^{(0)}(\cdot \mid \mathcal{D})$, is precisely the standard unscaled dimension, and the other orders can be more useful than it for certain complexity classes. To illustrate this, we mention some examples from circuit-size complexity. For a function $s : \mathbb{N} \to \mathbb{N}$, let $\mathrm{SIZE}(s(n))$ consist of all languages decidable by nonuniform Boolean circuit families of size $s(n)$. Lutz [12] showed that

$$\dim\left(\mathrm{SIZE}\left(\alpha \frac{2^n}{n}\right) \bigg| \mathrm{ESPACE}\right) = \alpha \tag{1.1}$$

for all $\alpha \in (0, 1)$. Circuit size bounds of the from $2^{\alpha n}$ and $2^{n^\alpha}$ are typically of more interest in complexity theory, but (1.1) implies that $\mathrm{SIZE}(2^{\alpha n})$ and $\mathrm{SIZE}(2^{n^\alpha})$ have dimension 0 in E for all $\alpha \in (0, 1)$. For these size bounds, the scaled dimensions are useful; in [9] it is shown that

$$\dim^{(1)}(\mathrm{SIZE}(2^{\alpha n}) \mid \mathrm{ESPACE}) = \alpha$$

and
$$\dim^{(2)}(\text{SIZE}(2^{n^\alpha}) \mid \text{ESPACE}) = \alpha$$

for any $\alpha \in (0, 1)$.

This paper uses scaled dimension to investigate polynomial-time spans and degrees and further understand the contrast between Theorems 1.1 and 1.2. We show that the same dichotomy also occurs between the $-3^{\text{rd}}$- and $-2^{\text{nd}}$-orders of scaled dimension. The main contribution of this paper is a strengthening of Theorem 1.1 to give a small span theorem for scaled dimension. (The following is a corollary of a stronger result proved in Theorem 6.3.)

**Theorem 1.3.** *Let $\mathcal{D} \in \{\text{E}, \text{EXP}, \text{ESPACE}, \text{EXPSPACE}\}$. For every $A \in \mathcal{D}$,*

$$\dim^{(-3)}(\text{P}_{\text{m}}(A) \mid \mathcal{D}) = 0$$

*or*

$$\dim^{(-3)}(\text{P}_{\text{m}}^{-1}(A) \mid \mathcal{D}) = 0.$$

*In particular, $\dim^{(-3)}(\deg_{\text{m}}^{\text{p}}(A) \mid \mathcal{D}) = 0$.*

In contrast, Theorem 1.2 is extended to scaled dimension at orders $i$ with $|i| \leq 2$.

**Theorem 1.4.** *Let $\mathcal{D} \in \{\text{E}, \text{EXP}, \text{ESPACE}, \text{EXPSPACE}\}$. For every $A \in \mathcal{D}$ and $-2 \leq i \leq 2$,*

$$\dim^{(i)}(\deg_{\text{m}}^{\text{p}}(A) \mid \mathcal{D}) = \dim^{(i)}(\text{P}_{\text{m}}(A) \mid \mathcal{D}).$$

This implies that Theorem 1.3 cannot be improved to $-2^{\text{nd}}$-order scaled dimension.

As an application of these results, we consider the scaled dimension of $\mathcal{C}_{\text{m}}^{\text{p}}(\text{E})$, the class of polynomial-time many-one complete sets for E, within ESPACE. Let $i \in \{-2, -1\}$. We extend a theorem of Lutz [13] to show that

$$\dim^{(i)}(\mathcal{C}_{\text{m}}^{\text{p}}(\text{E}) \mid \text{ESPACE}) > 0 \Rightarrow \text{P} = \text{BPP}.$$

On the other hand, we show that

$$\dim^{(i)}(\mathcal{C}_{\text{m}}^{\text{p}}(\text{E}) \mid \text{ESPACE}) < 1 \Rightarrow \text{P} \neq \text{PSPACE}.$$

Therefore, determining the $-1^{\text{st}}$ or $-2^{\text{nd}}$-order scaled dimension of $\mathcal{C}_{\text{m}}^{\text{p}}(\text{E})$ in ESPACE would derandomize BPP or separate P from PSPACE. In contrast, we also show that

$$\dim^{(-3)}(\mathcal{C}_{\text{m}}^{\text{p}}(\text{E}) \mid \text{ESPACE}) = 0$$

and

$$\dim^{(-2)}(\mathcal{C}_{\text{m}}^{\text{p}}(\text{E}) \mid \text{E}) = \dim^{(-1)}(\mathcal{C}_{\text{m}}^{\text{p}}(\text{E}) \mid \text{E}) = 1$$

hold without any hypothesis.

This paper is organized as follows. Section 2 contains the basic preliminaries and Section 3 reviews resource-bounded scaled dimension. We develop some new tools for computing scaled dimension in Section 4. The scaled dimensions of some auxiliary classes involving polynomial reductions are calculated in Section 5. Our small span theorem for scaled dimension is proved in Section 6. Section 7 shows that lower spans and degrees have the same dimension in orders $i$ with $-2 \leq i \leq 2$. Extensions of the results to $\leq_{1-\text{tt}}^{\text{P}}$-reductions are discussed in Section 8. The results on the scaled dimension of the complete sets for E are presented in Section 9. Section 10 concludes with a brief summary.

## 2 Preliminaries

The set of all finite binary strings is $\{0,1\}^*$. The empty string is denoted by $\lambda$ and $\{0,1\}^+ = \{0,1\}^* - \{\lambda\}$. We use the standard enumeration of binary strings $s_0 = \lambda, s_1 = 0, s_2 = 1, s_3 = 00, \ldots$. The length of a string $x \in \{0,1\}^*$ is denoted by $|x|$. We use the notation $\{0,1\}^{\leq n} = \{x \in \{0,1\}^* \mid |x| \leq n\}$ and $\{0,1\}^{>n} = \{x \in \{0,1\}^* \mid |x| > n\}$.

All *languages* (decision problems) in this paper are encoded as subsets of $\{0,1\}^*$. For a language $A \subseteq \{0,1\}^*$, we define $A_{\leq n} = \{x \in A \mid |x| \leq n\}$. We routinely identify $A$ with its infinite binary characteristic sequence according to the standard enumeration of binary strings. We write $A \upharpoonright n$ for the $n$-bit prefix of the characteristic sequence of $A$, and $A[n]$ for the $n^{\text{th}}$-bit of its characteristic sequence.

Let $\leq_r^{\text{P}}$ be a reducibility. For any $A \subseteq \{0,1\}^*$, let

$$\text{P}_r(A) = \{B \subseteq \{0,1\}^* \mid B \leq_r^{\text{P}} A\}$$

be the $\leq_r^{\text{P}}$-*lower span of* $A$,

$$\text{P}_r^{-1}(A) = \{B \subseteq \{0,1\}^* \mid A \leq_r^{\text{P}} B\}$$

be the $\leq_r^{\text{P}}$-*upper span of* $A$, and

$$\deg_r^{\text{p}}(A) = \text{P}_r(A) \cap \text{P}_r^{-1}(A)$$

be the $\leq_r^{\text{P}}$-*degree of* $A$. For any complexity class $\mathcal{D}$, the class of $\leq_r^{\text{P}}$-*hard languages for* $\mathcal{D}$ is

$$\mathcal{H}_r^{\text{p}}(\mathcal{D}) = \{A \subseteq \{0,1\}^* \mid \mathcal{D} \subseteq \text{P}_r(A)\},$$

and the class of $\leq_r^{\text{P}}$-*complete languages for* $\mathcal{D}$ is

$$\mathcal{C}_r^{\text{p}}(\mathcal{D}) = \mathcal{D} \cap \mathcal{H}_r^{\text{p}}(\mathcal{D}).$$

Let resource $\in \{\text{time}, \text{space}\}$ and let $t(n)$ be a resource bound. Let $l \in \mathbb{N}$. A function $f : \mathbb{N}^l \times \{0,1\}^* \to [0,\infty) \cap \mathbb{Q}$ is $t(n)$-*resource exactly computable* if there is a Turing machine that computes $f(k_1, \ldots, k_l, w)$ using at most $t(k_1 + \cdots + k_l + |w|)$ resource for all $(k_1, \ldots, k_l, w) \in \mathbb{N}^l \times \{0,1\}^*$. Let $g : \mathbb{N}^l \times \{0,1\}^* \to [0,\infty)$ be a real-valued function. An *approximation* of $g$ is a function $\hat{g} : \mathbb{N}^{l+1} \times \{0,1\}^* \to [0,\infty)$ such that

$$|g(x) - \hat{g}(r,x)| \leq 2^{-r}$$

for all $x \in \mathbb{N}^l \times \{0,1\}^*$ and $r \in \mathbb{N}$. We say that $g$ is $t(n)$-*resource computable* if there is an exactly $t(n)$-resource computable approximation $\hat{g}$ of $g$. A family of functions $(f_i : \mathbb{N}^l \times \{0,1\}^* \to [0,\infty) \mid i \in \mathbb{N})$ is *uniformly* $t(n)$-*resource (exactly) computable* if the function $f(i,x) = f_i(x)$ is $t(n)$-resource (exactly) computable.

A function $f$ is p-*computable* (respectively, pspace-*computable*) if it is $O(n^k)$-time (respectively, $O(n^k)$-space) computable for some $k \in \mathbb{N}$, and $f$ is p$_2$-*computable* (respectively, p$_2$space-*computable*) if it is $O(2^{(\log n)^k})$-time (respectively, $O(2^{(\log n)^k})$-space) computable for some $k \in \mathbb{N}$. Throughout this paper, unless otherwise specified, $\Delta$ denotes any of the *resource bounds* p, p$_2$, pspace, or p$_2$space. The concept of an *exactly* $\Delta$-*computable* function is defined analogously.

# 3 Scaled Dimension

Hitchcock, Lutz, and Mayordomo [9] introduced resource-bounded scaled dimension. This section briefly reviews the essentials of this theory.

The principle concept is a *scale*, which is a function $g : H \times [0, \infty) \to \mathbb{R}$, where $H = (a, \infty)$ for some $a \in \mathbb{R} \cup \{-\infty\}$. A scale must satisfy certain properties that are given in [9] and will not be discussed here. The canonical example of a scale is the function $g_0 : \mathbb{R} \times [0, \infty) \to \mathbb{R}$ defined by $g_0(m, s) = sm$. This scale is used in the standard (unscaled) dimension. Other scales of interest are obtained from $g_0$ by rescaling and reflection operations.

**Definition.** *Let $g : H \times [0, \infty) \to \mathbb{R}$ be a scale.*

1. *The* first rescaling *of $g$ is the scale $g^{\#} : H^{\#} \times [0, \infty) \longrightarrow \mathbb{R}$ defined by*

$$H^{\#} = \{2^m \mid m \in H\}$$

$$g^{\#}(m, s) = 2^{g(\log m, s)}.$$

2. *The* reflection *of $g$ is the scale $g^R : H \times [0, \infty) \to \mathbb{R}$ defined by*

$$g^R(m, s) = \begin{cases} m + g(m, 0) - g(m, 1 - s) & \text{if } 0 \leq s \leq 1 \\ g(m, s) & \text{if } s \geq 1. \end{cases}$$

A family of scales, one for each integers, is defined as follows.

**Definition.** 1. *For each $i \in \mathbb{N}$, define $a_i$ by the recurrence $a_0 = -\infty, a_{i+1} = 2^{a_i}$.*

2. *For each $i \in \mathbb{Z}$, define the $i^{\text{th}}$ scale $g_i : (a_{|i|}, \infty) \times [0, \infty) \to \mathbb{R}$ by the following recursion.*

   (a) $g_0(m, s) = sm$.
   (b) *For $i \geq 0$, $g_{i+1} = g_i^{\#}$.*
   (c) *For $i < 0$, $g_i = g_{-i}^R$.*

For clarity, we compute the first few scales. For all $s \in [0, 1]$, if $m > a_{|i|}$, then $g_i(m, s)$ is defined by

$$\begin{aligned}
g_3(m, s) &= 2^{2^{(\log \log m)^s}} \\
g_2(m, s) &= 2^{(\log m)^s} \\
g_1(m, s) &= m^s \\
g_0(m, s) &= sm \\
g_{-1}(m, s) &= m + 1 - m^{1-s} \\
g_{-2}(m, s) &= m + 2 - 2^{(\log m)^{1-s}} \\
g_{-3}(m, s) &= m + 4 - 2^{2^{(\log \log m)^{1-s}}}.
\end{aligned}$$

Scaled dimension is defined using functions called *scaled gales*. The more familiar concepts of *gales* [12] and *martingales* [14] are special cases in the following definition.

**Definition.** *Let $i \in \mathbb{Z}$ and let $s \in [0, \infty)$.*

1. A $i^{\text{th}}$-order scaled $s$-gale (briefly, an $s^{(i)}$-gale) is a function $d : \{0,1\}^{>a_{|i|}} \to [0, \infty)$ such that for all $w \in \{0,1\}^*$ with $|w| > a_{|i|}$,

$$d(w) = 2^{-\Delta g_i(|w|,s)}[d(w0) + d(w1)], \qquad (3.1)$$

where $\Delta g_i : (a_{|i|}, \infty) \times [0, \infty) \to \mathbb{R}$ is defined by

$$\Delta g_i(m, s) = g_i(m+1, s) - g_i(m, s).$$

2. An $s$-gale is an $s^{(0)}$-gale, that is, a function $d : \{0,1\}^* \to [0, \infty)$ satisfying

$$d(w) = 2^{-s}[d(w0) + d(w1)]$$

for all $w \in \{0,1\}^*$.

3. A martingale is a 1-gale, that is, a function $d : \{0,1\}^* \to [0, \infty)$ satisfying

$$d(w) = \frac{d(w0) + d(w1)}{2}$$

for all $w \in \{0,1\}^*$.

*Success sets* are a crucial concept for resource-bounded measure, and also for scaled dimension.

**Definition.** Let $d : \{0,1\}^{>a} \to [0, \infty)$, where $a \in \mathbb{Z}$.

1. We say that $d$ succeeds *on a language* $A \subseteq \{0,1\}^*$ if

$$\limsup_{n \to \infty} d(A \restriction n) = \infty.$$

2. The success set *of $d$ is*

$$S^\infty[d] = \{A \subseteq \{0,1\}^* \mid d \text{ succeeds on } A\}.$$

Resource-bounded measure is defined using success sets of martingales. Here $\Delta$ denotes any of the resource bounds $\{\text{p}, \text{p}_2, \text{pspace}, \text{p}_2\text{space}\}$, and $R(\Delta)$ is the following exponential-time or -space complexity class.

$$\begin{array}{rcl}
R(\text{p}) & = & \text{E} & = & \text{DTIME}(2^{O(n)}) \\
R(\text{p}_2) & = & \text{EXP} & = & \text{DTIME}(2^{n^{O(1)}}) \\
R(\text{pspace}) & = & \text{ESPACE} & = & \text{DSPACE}(2^{O(n)}) \\
R(\text{p}_2\text{space}) & = & \text{EXPSPACE} & = & \text{DSPACE}(2^{n^{O(1)}})
\end{array}$$

**Definition.** Let $\mathcal{C}$ be a class of languages.

1. We say that $\mathcal{C}$ has $\Delta$-measure 0, and write $\mu_\Delta(\mathcal{C}) = 0$, if there is a $\Delta$-computable martingale $d$ such that $\mathcal{C} \subseteq S^\infty[d]$.

2. We say that $\mathcal{C}$ has measure 0 in $R(\Delta)$, and write $\mu(\mathcal{C} \mid R(\Delta)) = 0$, if $\mu_\Delta(\mathcal{C} \cap R(\Delta)) = 0$.

The *measure conservation theorem* of Lutz [14] asserts that $\mu_\Delta(R(\Delta)) \neq 0$, justifying the definition of measure in $R(\Delta)$ above.

Success sets of scaled gales are used to define scaled dimension.

**Definition.** Let $\mathcal{C}$ be a class of languages and $i \in \mathbb{Z}$.

1. The $i^{\text{th}}$-order scaled $\Delta$-dimension of $\mathcal{C}$ is

$$\dim_{\Delta}^{(i)}(\mathcal{C}) = \inf \left\{ s \,\middle|\, \begin{array}{l} \text{there exists a } \Delta\text{-computable} \\ s^{(i)}\text{-scaled gale } d \text{ for which } \mathcal{C} \subseteq S^{\infty}[d] \end{array} \right\}.$$

2. The $i^{\text{th}}$-order scaled dimension of $\mathcal{C}$ within $R(\Delta)$ is

$$\dim^{(i)}(\mathcal{C} \mid R(\Delta)) = \dim_{\Delta}^{(i)}(\mathcal{C} \cap R(\Delta)).$$

The $0^{\text{th}}$-order dimension $\dim_{\Delta}^{(0)}(\cdot)$ is precisely the dimension $\dim_{\Delta}(\cdot)$ of Lutz [12], and the other orders are interpreted as rescalings of this concept.

The following lemma relates resource-bounded scaled dimension to resource-bounded measure.

**Lemma 3.1.** ([9]) *For any class $\mathcal{C}$ of languages and $i \in \mathbb{Z}$,*

$$\dim_{\Delta}^{(i)}(\mathcal{C}) < 1 \Rightarrow \mu_{\Delta}(\mathcal{C}) = 0$$

*and*

$$\dim^{(i)}(\mathcal{C} \mid R(\Delta)) < 1 \Rightarrow \mu(\mathcal{C} \mid R(\Delta)) = 0.$$

The following is another key property of scaled dimension.

**Theorem 3.2.** ([9]) *Let $\mathcal{C}$ be a class of languages and $i \in \mathbb{Z}$. If $\dim_{\Delta}^{(i+1)}(\mathcal{C}) < 1$, then $\dim_{\Delta}^{(i)}(\mathcal{C}) = 0$.*

This theorem tells us that for every class $\mathcal{C}$, the sequence of dimensions $\dim_{\Delta}^{(i)}(X)$ for $i \in \mathbb{Z}$ satisfies exactly one of the following three conditions.

(i) $\dim_{\Delta}^{(i)}(\mathcal{C}) = 0$ for all $i \in \mathbb{Z}$.

(ii) $\dim_{\Delta}^{(i)}(\mathcal{C}) = 1$ for all $i \in \mathbb{Z}$.

(iii) There exist $i^* \in \mathbb{Z}$ such that $\dim_{\Delta}^{(i)}(\mathcal{C}) = 0$ for all $i < i^*$ and $\dim_{\Delta}^{(i)}(\mathcal{C}) = 1$ for all $i > i^*$.

## 4 Measures, Log-Loss, and Scaled Dimension

This section provides some tools involving measures and the log-loss concept that are useful for working with the scaled dimensions. It was shown in [8] that *log-loss unpredictability* is equivalent to dimension. We similarly characterize scaled dimension using the log-loss of measures.

**Definition.** *A* measure *is a function $\rho : \{0,1\}^* \to [0,\infty)$ satisfying*

$$\rho(w) = \rho(w0) + \rho(w1)$$

*for all $w \in \{0,1\}^*$.*

Measures have the following fundamental relationship with scaled gales.

**Observation 4.1.** *Let $i \in \mathbb{Z}$ and $s \in [0, \infty)$.*

1. If $\rho : \{0,1\}^* \to [0,\infty)$ is a measure, then the function $d_\rho : \{0,1\}^{>a_{|i|}} \to [0,\infty)$ defined by

$$d_\rho(w) = 2^{g_i(|w|,s)} \rho(w)$$

for all $w \in \{0,1\}^{>a_{|i|}}$ is an $s^{(i)}$-gale.

2. If $d : \{0,1\}^{>a_{|i|}} \to [0,\infty)$ is an $s^{(i)}$-gale, then the function $\rho_d : \{0,1\}^* : [0,\infty)$ defined by

$$\rho_d(w) = 2^{-g_i(|w|,s)} d(w)$$

for all $w \in \{0,1\}^{>a_{|i|}}$ and

$$\rho_d(w) = \sum_{|v|=a_{|i|}+1-|w|} \rho_d(wv)$$

for all $w \in \{0,1\}^{\leq a_{|i|}}$ is a measure.

The following lemma relates the scaled dimension of a class to limits involving scales and logarithms of measures.

**Lemma 4.2.** *Let $\mathcal{C}$ be a class of languages and let $i \in \mathbb{Z}$.*

1. *If $s > \dim_\Delta^{(i)}(\mathcal{C})$, then there is a $\Delta$-computable measure $\rho$ such that*

$$\limsup_{n\to\infty} g_i(n,s) + \log \rho(A \upharpoonright n) = \infty$$

*for all $A \in \mathcal{C}$.*

2. *If $s < \dim_\Delta^{(i)}(\mathcal{C})$, then for any $\Delta$-computable measure $\rho$ there is an $A_\rho \in \mathcal{C}$ such that*

$$\lim_{n\to\infty} g_i(n,s) + \log \rho(A_\rho \upharpoonright n) = -\infty.$$

*Proof.* Let $r$ be rational with $s > r > \dim_\Delta^{(i)}(\mathcal{C})$ and let $d$ be a $\Delta$-computable $r^{(i)}$-gale succeeding on $\mathcal{C}$. Then the measure $\rho_d$ from Observation 4.1 is also $\Delta$-computable. Let $A \in \mathcal{C}$. There are infinitely many $n \in \mathbb{N}$ such that $d(A \upharpoonright n) \geq 1$ since $A \in S^\infty[d]$. For such $n$,

$$\begin{aligned} g_i(n,s) + \log \rho_d(A \upharpoonright n) &= g_i(n,s) - g_i(n,r) + \log d(A \upharpoonright n) \\ &\geq g_i(n,s) - g_i(n,r). \end{aligned}$$

Part 1 follows because $r < s$.

For part 2, let $\rho$ be a $\Delta$-computable measure. Let $t$ be rational with $s < t < \dim_\Delta^{(i)}(\mathcal{C})$ and obtain the $t^{(i)}$-gale $d_\rho$ from Observation 4.1. Then $\mathcal{C} \not\subseteq S^\infty[d_\rho]$ because $d_\rho$ is $\Delta$-computable, so there is an $A_\rho \in \mathcal{C}$ and a constant $c$ such that $d(A \upharpoonright n) \leq c$ for all $n > a_{|i|}$. Then

$$\begin{aligned} g_i(n,s) + \log \rho(A \upharpoonright n) &= g_i(n,s) - g_i(n,t) + \log d_\rho(A \upharpoonright n) \\ &\leq g_i(n,s) - g_i(n,t) + \log c, \end{aligned}$$

so the claim follows because $s < t$. $\square$

Lemma 4.2 asserts that if the $i^{\text{th}}$-order scaled dimension of a class $\mathcal{C}$ is less than $s$ then there is a measure $\rho$ such that for every $A \in \mathcal{C}$, there are prefixes $w \sqsubseteq A$ where the *log-loss* quantity

$$-\log \rho(w)$$

is arbitrarily less than $g_i(|w|, s)$.

It is often convenient to replace computable measures by exactly computable measures.

**Lemma 4.3.** *Let $\rho$ be a measure that is computable in $t(n)$ time (respectively, space). Then there is a measure $\tilde{\rho}$ that is exactly computable in $O(n \cdot t(3n))$ time (respectively, space) such that*

$$\log \tilde{\rho}(w) \geq \log \rho(w) - c$$

*for all $w \in \{0,1\}^*$ where $c$ is a constant that is independent of $w$.*

*Proof.* We assume that $\rho(w) \geq 2^{-|w|}$ for all $w \in \{0,1\}^*$. (If $\rho$ does not satisfy this condition, add $2^{-|w|}$ to $\rho(w)$ to obtain a measure that does and use it instead.)

Let $\hat{\rho} : \mathbb{N} \times \{0,1\}^* \to [0, \infty)$ be an approximation of $\rho$. For all $w \in \{0,1\}^*$, define

$$\rho'(w) = \hat{\rho}(2|w|, w).$$

The measure $\tilde{\rho} : \{0,1\}^* \to [0, \infty)$ is defined by

$$\tilde{\rho}(\lambda) = \rho'(0) + \rho'(1)$$

and

$$\tilde{\rho}(wb) = \frac{\rho'(wb)}{\rho'(w0) + \rho'(w1)} \tilde{\rho}(w)$$

for all $w \in \{0,1\}^*$ and $b \in \{0,1\}$. If $\hat{\rho}$ is exactly computable in $t(n)$ time, then $\rho'(w)$ is computable in $t(2|w| + |w|)$ time, so we can exactly compute $\tilde{\rho}(w)$ in $O(|w| \cdot t(3|w|))$ time. Similarly, if $\hat{\rho}$ is computable in $t(n)$ space, then $\tilde{\rho}$ is computable in $O(n \cdot t(3n))$ space.

For any $w \in \{0,1\}^+$, we have

$$\begin{aligned}
\frac{\rho'(w)}{\rho'(w0) + \rho'(w1)} &\geq \frac{\rho(w) - 2^{-2|w|}}{\rho(w0) + 2^{-2(|w|+1)} + \rho(w1) + 2^{-2(|w|+1)}} \\
&= \frac{\rho(w) - 2^{-2|w|}}{\rho(w) + 2^{-2|w|-1}} \\
&\geq \frac{2^{-|w|} - 2^{-2|w|}}{2^{-|w|} + 2^{-2|w|-1}} \\
&= \frac{2^{|w|} - 1}{2^{|w|} + \frac{1}{2}} \\
&= 1 - \frac{\frac{3}{2}}{2^{|w|} + \frac{1}{2}} \\
&\geq 1 - \frac{2}{2^{|w|+1}} \\
&= 1 - 2^{-|w|},
\end{aligned}$$

with the second inequality holding because

$$\frac{\alpha - \epsilon}{\alpha + \frac{\epsilon}{2}}$$

is an increasing function of $\alpha$ and $2^{-|w|}$ is the minimum possible value for $\rho(w)$. Therefore

$$\begin{aligned}
\log \tilde{\rho}(w) &= \log \left( \prod_{i=1}^{|w|} \frac{\rho'(w \upharpoonright i)}{\rho'((w \upharpoonright i - 1)0) + \rho'((w \upharpoonright i - 1)1)} \right) \tilde{\rho}(\lambda) \\
&= \log \rho'(w) + \sum_{i=1}^{|w|-1} \log \frac{\rho'(w \upharpoonright i)}{\rho'((w \upharpoonright i)0) + \rho'((w \upharpoonright i)1)} + \log \frac{\rho'(0) + \rho'(1)}{\rho'(0) + \rho'(1)} \\
&\geq \log \rho'(w) + \sum_{i=1}^{|w|-1} \log 1 - 2^{-i} \\
&\geq \log \rho'(w) + \sum_{i=1}^{|w|-1} \frac{-2^{-i}}{\ln 2} \\
&\geq \log \rho'(w) - \frac{1}{\ln 2}.
\end{aligned}$$

Also, again using the fact that $\rho(w) \geq 2^{-|w|}$,

$$\begin{aligned}
\log \rho'(w) &\geq \log \left[ \rho(w) - 2^{-2|w|} \right] \\
&\geq \log \rho(w) \frac{2^{-|w|} - 2^{-2|w|}}{2^{-|w|}} \\
&= \log \rho(w)(1 - 2^{-|w|}) \\
&\geq \log \rho(w) - \frac{2^{-|w|}}{\ln 2}.
\end{aligned}$$

Combining the above, we have

$$\log \tilde{\rho}(w) \geq \log \rho(w) - \frac{2}{\ln 2}$$

for all $w \in \{0,1\}^+$. The lemma holds with $c = \left\lceil \max \left\{ \frac{2}{\ln 2}, \log \frac{\rho(\lambda)}{\tilde{\rho}(\lambda)} \right\} \right\rceil$. □

The measures that are exactly computable within a fixed time or space bound are uniformly exactly computable with slightly more time or space.

**Lemma 4.4.** *For any time constructible functions $t(n)$ and $t'(n)$ with $t(n) \log t(n) = o(t'(n))$, the family of exactly $t(n)$-time computable measures is uniformly exactly computable in $t'(n)$-time. If $t(n) = o(t'(n))$, then the family of exactly $t(n)$-space computable measures is uniformly exactly computable in $t'(n)$-space.*

*Proof.* There is a uniform enumeration $(M_i \mid i \in \mathbb{N})$ of all $t(n)$-time (respectively, $t(n)$-space) clocked Turing machines such that for all $i \in \mathbb{N}$, $M_i(w)$ can be computed in $O(t(|w|) \log t(|w|))$ time (respectively, $O(t(|w|))$ space) for all $w \in \{0,1\}^*$. (Here the constants in the $O(\cdot)$ depend on $i$ but not on $|w|$.) Define $\rho_i : \{0,1\}^* \to [0, \infty)$ inductively by $\rho_i(\lambda) = M_i(\lambda)$ and

$$\rho_i(w0) = \begin{cases} M_i(w0) & \text{if } M_i(w0) \leq \rho_i(w) \\ M_i(w) & \text{otherwise,} \end{cases}$$

$$\rho_i(w1) = \rho_i(w) - \rho_i(w0)$$

for all $w \in \{0,1\}^*$. Then each $\rho_i$ is a measure, and the family is uniformly computable in $t'(n)$ time (respectively, $t'(n)$ space). Also, if $\rho$ is a measure that is exactly computed by $M_i$ in $t(n)$ time, then $\rho_i(w) = \rho(w)$ for all $w$. □

Uniformly exactly computable families of measures can be combined into a single measure in an efficient manner.

**Lemma 4.5.** *Let $(\rho_k \mid k \in \mathbb{N})$ be a uniformly exactly $\Delta$-computable family of measures. There is a $\Delta$-computable measure $\rho^*$ such that for any $k$, there is a constant $c_k$ such that*

$$\log \rho^*(w) \geq \log \rho_k(w) - c_k$$

*for all $w \in \{0,1\}^*$.*

*Proof.* Define

$$\rho^*(w) = \sum_{k=0}^{\infty} \frac{\rho_k(w)}{2^k \rho_k(\lambda)}.$$

Then $\rho$ is a measure by linearity. Also, $\rho^*$ is $\Delta$-computable by the approximation function $\hat{\rho}^* : \mathbb{N} \times \{0,1\}^* \to [0, \infty)$ defined by

$$\hat{\rho}^*(r, w) = \sum_{k=0}^{r} \frac{\rho_k(w)}{2^k \rho_k(\lambda)}$$

since

$$\begin{aligned} \left| \rho^*(w) - \hat{\rho}^*(r, w) \right| &= \sum_{k=r+1}^{\infty} \frac{\rho_k(w)}{2^k \rho_k(\lambda)} \\ &\leq \sum_{k=r+1}^{\infty} \frac{\rho_k(\lambda)}{2^k \rho_k(\lambda)} \\ &= 2^{-r}. \end{aligned}$$

Let $k \in \mathbb{N}$. For any $w \in \{0,1\}^*$,

$$\begin{aligned} \log \rho^*(w) &\geq \log \frac{\rho_k(w)}{2^k \rho_k(\lambda)} \\ &= \log \rho_k(w) - k - \rho_k(\lambda), \end{aligned}$$

so the lemma holds with $c_k = k + \rho_k(\lambda)$. □

We now combine the preceding lemmas to obtain a tool that will be useful in calculating scaled dimensions.

**Theorem 4.6.** *Let $\mathcal{C}$ be a class of languages, $i \in \mathbb{Z}$, and $k \in \mathbb{N}$.*

1. *If for each $A \in \mathcal{C}$ there is a measure $\rho_A$ computable in $O(n^k)$ time such that*

$$(\exists c_A \in \mathbb{Z})(\exists^\infty n) g_i(n,s) + \log \rho_A(A \upharpoonright n) \geq c_A, \tag{4.1}$$

   *then $\dim_{\mathrm{p}}^{(i)}(\mathcal{C}) \leq s$.*

2. *If for each $A \in \mathcal{C}$ there is a measure $\rho_A$ computable in $O(2^{\log n^k})$ time such that (4.1) holds, then then $\dim_{\mathrm{p2}}^{(i)}(\mathcal{C}) \leq s$.*

3. *If for each $A \in \mathcal{C}$ there is a measure $\rho_A$ computable in $O(n^k)$ space such that (4.1) holds, then then $\dim_{\mathrm{pspace}}^{(i)}(\mathcal{C}) \leq s$.*

4. *If for each $A \in \mathcal{C}$ there is a measure $\rho_A$ computable in $O(2^{\log n^k})$ space such that (4.1) holds, then then $\dim_{\mathrm{p2space}}^{(i)}(\mathcal{C}) \leq s$.*

*Proof.* From Lemmas 4.3, 4.4, and 4.5 we obtain an exactly $\Delta$-computable measure $\rho$ such that $\log \rho(w) \geq \log \rho_A(w) - b_A$ for all $w \in \{0,1\}^*$ where $b_A$ is a constant that depends on $A$ but not on $w$.

Let $t > s$. For any $A \in \mathcal{C}$,

$$g_i(n,t) + \log \rho(A \upharpoonright n) \geq g_i(n,t) - g_i(n,s) + c_A - b_A$$

for infinitely many $n$. Therefore

$$\limsup_{n \to \infty} g_i(n,t) + \log \rho(A \upharpoonright n) = \infty$$

since $t > s$. It follows from the converse of Lemma 4.2(2) that $\dim_\Delta(\mathcal{C}) \leq t$. $\square$

## 5  Scaled Non-Bi-Immunity and Compressibility

In this section we introduce some classes involving scales, non-bi-immunity, and compressibility by polynomial-time reductions and calculate their scaled dimensions.

A Turing machine $M$ is *consistent* with a language $A \subseteq \{0,1\}^*$ if for all $x \in \{0,1\}^*$,

$$M(x) \text{ halts} \iff M(x) = A(x).$$

Let $t$ be a time bound. The *fast set of $M$ with respect to $t$* is

$$F_M^t = \{x \in \{0,1\}^* \mid \mathrm{time}_M(x) \leq t(|x|)\}.$$

Recall that $A$ is *not* $\mathrm{DTIME}(t)$-*bi-immune* if there is a machine $M$ consistent with $A$ such that $F_M^t$ is infinite.

**Definition.** *For any time bound $t$, let $X(t)$ be the class of all languages that are not $\mathrm{DTIME}(t)$-bi-immune.*

Let $A \subseteq \{0,1\}^*$ and $f : \{0,1\}^* \to \{0,1\}^*$. We say that $f$ is a *many-one reduction of $A$* if there is some $B \subseteq \{0,1\}^*$ such that $x \in A \iff f(x) \in B$. The *collision set* of $f$ is

$$C_f = \{s_i | (\exists j < i) f(s_i) = f(s_j)\}.$$

Recall that $A$ is *compressible by $\leq_{\mathrm{m}}^{\mathrm{DTIME}(t)}$-reductions* if there exists an $f \in \mathrm{DTIMEF}(t)$ that is a many-one reduction of $A$ and has $C_f$ infinite [10].

**Definition.** *For any time bound $t$, let $C(t)$ be the class of all languages that are compressible by $\leq_{\mathrm{m}}^{\mathrm{DTIME}(t)}$-reductions.*

The following theorem asserts that almost every language in E is $\mathrm{DTIME}(2^{cn})$-bi-immune [18] and incompressible by $\leq_{\mathrm{m}}^{\mathrm{DTIME}(2^{cn})}$-reductions [10].

**Theorem 5.1.** (Mayordomo [18], Juedes and Lutz [10]) *For all $c \in \mathbb{N}$,*

$$\mu_{\mathrm{p}}(X(2^{cn})) = \mu_{\mathrm{p}}(C(2^{cn})) = 0$$

*and*

$$\mu_{\mathrm{p}_2}(X(2^{n^c})) = \mu_{\mathrm{p}_2}(C(2^{n^c})) = 0.$$

The next two definitions introduce scaled versions of $X(t)$ and $C(t)$.

**Definition.** *For any $i \in \mathbb{Z}$, $\alpha \in [0,1]$, and time bound $t$, let*

$$X_\alpha^{(i)}(t) = \left\{ A \subseteq \{0,1\}^* \;\middle|\; \begin{array}{l} (\exists M) M \text{ is consistent with } A \text{ and} \\ (\exists^\infty n) \#(1, F_M^t \upharpoonright n) \geq n - g_i(n, \alpha) \end{array} \right\}.$$

That is, $X_\alpha^{(i)}(t)$ consists of the languages that are not $\mathrm{DTIME}(t)$-bi-immune in a particular strong way: for infinitely many $n$, all but $g_i(n, \alpha)$ of the first $n$ strings can be decided in less than $t$ time by a consistent Turing machine.

**Definition.** *For any $i \in \mathbb{Z}$, $\alpha \in [0,1]$, and time bound $t$, let*

$$C_\alpha^{(i)}(t) = \left\{ A \in \{0,1\}^* \;\middle|\; \begin{array}{l} (\exists f \in \mathrm{DTIMEF}(t)) \; f \text{ is a many-one reduction of } A \\ \text{and } (\exists^\infty n) \#(1, C_f \upharpoonright n) \geq n - g_i(n, \alpha) \end{array} \right\}.$$

In other words, $C_\alpha^{(i)}(t)$ is the class of languages compressible by $\leq_{\mathrm{m}}^{\mathrm{DTIME}(t)}$-reductions where for infinitely many $n$, all but $g_i(n, \alpha)$ of the first $n$ strings have collisions under some reduction.

For $\alpha < 1$, $X_\alpha^{(i)}(2^n) \subseteq X(2^n)$ and $C_\alpha^{(i)}(2^n) \subseteq C(2^n)$, so Theorem 5.1 implies that $X_\alpha^{(i)}(2^n)$ and $C_\alpha^{(i)}(2^n)$ have measure 0. We now refine this by calculating their scaled dimensions.

**Theorem 5.2.** *For all $i \in \mathbb{Z}$, $c \geq 1$, and $\alpha \in [0,1]$,*

$$\dim_{\mathrm{p}}^{(i)}(X_\alpha^{(i)}(2^{cn})) = \dim_{\mathrm{p}}^{(i)}(C_\alpha^{(i)}(2^{cn})) = \alpha$$

*and*

$$\dim_{\mathrm{p}_2}^{(i)}(X_\alpha^{(i)}(2^{n^c})) = \dim_{\mathrm{p}_2}^{(i)}(C_\alpha^{(i)}(2^{n^c})) = \alpha.$$

*Proof.* We focus on the p-dimension portion of the theorem; the argument for $p_2$-dimension is identical. If $\alpha = 0$, this is trivial, so assume $\alpha \in (0, 1]$. Let $s, t > 0$ be arbitrary rationals with $s < \alpha < t$. It suffices to show that

$$s \leq \dim_p^{(i)}(X_\alpha^{(i)}(2^{cn})) \leq \dim_p^{(i)}(C_\alpha^{(i)}(2^{cn})) \leq t.$$

The inequality $\dim_p^{(i)}(X_\alpha^{(i)}(2^{cn})) \leq \dim_p^{(i)}(C_\alpha^{(i)}(2^{cn}))$ holds because of the inclusion $X_\alpha^{(i)}(2^{cn}) \subseteq C_\alpha^{(i)}(2^{cn})$.

For the lower bound, let $\rho$ be any p-computable measure, say computable in $n^k$ time. We define a language $A$ inductively by lengths. Let $s' \in (s, \alpha)$ be rational. The first $\lfloor g_i(2^n, s') \rfloor$ bits of $A_{=n}$ are set by diagonalization to minimize $\rho$. The remaining $2^n - \lfloor g_i(2^n, s') \rfloor$ bits are identically 0. More formally, if $x$ is the characteristic string of $A_{\leq n-1}$, we choose $v \in \{0,1\}^{\lfloor g_i(2^n, s') \rfloor}$ so that $\rho(xv)$ is minimized, and let $A_{=n}$ have characteristic string $v0^{2^n - \lfloor g_i(2^n, s') \rfloor}$. Then $A$ is easily in $X_{s'}^{(i)}(2^{cn}) \subseteq X_\alpha^{(i)}(2^{cn})$. Let $w \sqsubseteq A$, and let $n$ be such that $2^n - 1 \leq |w| < 2^{n+1} - 1$. Then if $|w| \leq 2^n - 1 + \lfloor g_i(2^n, s') \rfloor$, we have

$$\rho(w) \leq \rho(w \restriction 2^n - 1) 2^{-(|w| - (2^n - 1))},$$

and if $|w| \geq 2^n - 1 + \lfloor g_i(2^n, s') \rfloor$, we have

$$\begin{aligned}
\rho(w) &\leq \rho(w \restriction (2^n - 1 + \lfloor g_i(2^n, s') \rfloor)) \\
&\leq \frac{\rho(w \restriction 2^n - 1)}{2^{2^n - 1 + \lfloor g_i(2^n, s') \rfloor - (2^n - 1)}} \\
&= \frac{\rho(w \restriction 2^n - 1)}{2^{\lfloor g_i(2^n, s') \rfloor}}.
\end{aligned}$$

Therefore, in either case,

$$\begin{aligned}
\log \rho(w) + g_i(|w|, s) &\leq \log \rho(w \restriction 2^n - 1) - \lfloor g_i(2^n, s') \rfloor + g_i(|w|, s) \\
&\leq \log \rho(w \restriction 2^n - 1) + g_i(2^{n+1} - 1, s) - g_i(2^n, s').
\end{aligned}$$

As $g_i(2^n, s') - g_i(2^{n+1} - 1, s) \to \infty$ as $n \to \infty$ since $s < s'$, it follows that

$$\lim_{n \to \infty} \log \rho(A \restriction n) + g_i(n, s) = -\infty.$$

Since $\rho$ is an arbitrary p-computable measure, the converse of Lemma 4.2(1) implies that $\dim_p^{(i)}(X_\alpha^{(i)}) \geq s$.

Now we prove the upper bound. Let $A \in C_\alpha^{(i)}(2^{cn})$ by a function $f \in \text{DTIMEF}(2^{cn})$. Define a measure $\rho$ inductively by $\rho(\lambda) = 1$ and for all $w \in \{0,1\}^*, b \in \{0,1\}$,

1. If $f(s_i) \neq f(s_{|w|})$ for all $i < |w|$, then

$$\rho(wb) = \frac{\rho(w)}{2}.$$

2. Otherwise, let $i = \min\{i < |w| \mid f(s_i) = f(s_{|w|})\}$ and define

$$\rho(wb) = \begin{cases} \rho(w) & \text{if } b = w[i] \\ 0 & \text{if } b \neq w[i]. \end{cases}$$

Then for all $w \sqsubseteq A$,

$$\begin{aligned} \log \rho(w) &= -\#(0, C_f \upharpoonright |w|) \\ &= \#(1, C_f \upharpoonright |w|) - |w|. \end{aligned}$$

Whenever $\#(1, C_f \upharpoonright n) \geq n - g_i(n, \alpha)$, we have

$$\begin{aligned} g_i(n,t) + \log \rho(A \upharpoonright n) &= g_i(n,t) + \#(1, C_f \upharpoonright n) - n \\ &\geq g_i(n,t) - g_i(n,\alpha). \end{aligned}$$

This happens infinitely often, so

$$\limsup_{n \to \infty} g_i(n,t) + \log \rho(A \upharpoonright n) = \infty$$

because $t > \alpha$. Also, $\rho$ is computable in $O(|w| \cdot 2^{c \log |w|}) = O(|w|^{c+1})$ time. Such a $\rho$ can be defined for each $A \in C_\alpha^{(i)}(2^{cn})$, so $\dim_\mathrm{p}^{(i)}(C_\alpha^{(i)}(2^{cn})) \leq t$ follows by Theorem 4.6. $\square$

## 6 Small Span Theorem

In this section we establish our small span theorem for scaled dimension. We begin with a simple, but important, lemma about the scales.

**Lemma 6.1.** *For all $k \geq 1$ and $s, t \in (0, 1)$, $g_3(2^{n^k}, s) = o(g_2(2^n, t))$.*

*Proof.* We have

$$g_3(2^{n^k}, s) = 2^{2^{\left(\log \log 2^{n^k}\right)^s}} = 2^{2^{(k \log n)^s}}$$

and

$$g_2(2^n, t) = 2^{(\log 2^n)^t} = 2^{n^t} = 2^{2^{t \log n}}.$$

The lemma holds since $(k \log n)^s = o(t \log n)$. $\square$

Juedes and Lutz [10] proved that the upper spans of incompressible languages are small. Specifically, for any language $A \in \mathrm{EXP}$ that is incompressible by $\leq_\mathrm{m}^\mathrm{P}$-reductions, they showed that $\mu_{\mathrm{p}_2}(\mathrm{P}_\mathrm{m}^{-1}(A)) = 0$, and if additionally $A \in \mathrm{E}$, then $\mu_\mathrm{p}(\mathrm{P}_\mathrm{m}^{-1}(A)) = 0$. The following theorem is a scaled dimension analogue of this. For any $i \in \mathbb{Z}$, let

$$C_\alpha^{(i)}(\mathrm{poly}) = \bigcup_{c \in \mathbb{N}} C_\alpha^{(i)}(n^c + c).$$

**Theorem 6.2.** *Let $\alpha \in (0, 1)$.*

1. *Let $\Delta \in \{\mathrm{p}, \mathrm{pspace}\}$. For any $B \in R(\Delta) - C_\alpha^{(1)}(\mathrm{poly})$, $\dim_\Delta^{(-3)}(\mathrm{P}_\mathrm{m}^{-1}(B)) = 0$.*

2. *Let $\Delta \in \{\mathrm{p}_2, \mathrm{p}_2\mathrm{space}\}$. For any $B \in R(\Delta) - C_\alpha^{(2)}(\mathrm{poly})$, $\dim_\Delta^{(-3)}(\mathrm{P}_\mathrm{m}^{-1}(B)) = 0$.*

*Proof.* We first give the proof for $\Delta = \mathrm{p}$. Let $B \in \mathrm{E} - C_\alpha^{(1)}(\mathrm{poly})$ and let $M$ be a Turing machine that decides $B$ in $O(2^{cn})$ time. Assume $B \leq_\mathrm{m}^\mathrm{P} C$ via $f$ where $f$ is computable in $n^k$ time almost everywhere. Then for all sufficiently large $n$,

$$f(B_{\leq n}) \subseteq C_{\leq n^k} \tag{6.1}$$

and

$$|f(B_{\leq n})| \geq g_1(2^{n+1} - 1, \alpha) \geq g_1(2^n, \alpha), \tag{6.2}$$

with the latter holding because $B \notin C_\alpha^{(1)}(\mathrm{poly})$.

Let $r \in \mathbb{N}$ such that $\frac{1}{r} < \alpha$. Define $d : \mathbb{N} \to \mathbb{N}$ by $d(n) = \lfloor n/r \rfloor$. For each $n \in \mathbb{N}$ we define a measure $\rho_n : \{0,1\}^* \to [0,1]$ by

$$\rho_n(\lambda) = 2^{-n}$$

and for all $w \in \{0,1\}^*$ and $b \in \{0,1\}$,

1. If $|w| < 2^{d(n)}$ or $\left[(\forall i < 2^{n+1} - 1) f(s_i) \neq f(s_{|w|})\right]$, then

$$\rho_n(wb) = \frac{\rho_n(w)}{2}.$$

2. Otherwise, let $i = \min\left\{i < 2^{n+1} - 1 \,\big|\, f(s_i) = f(s_{|w|})\right\}$ and define

$$\rho_n(wb) = \begin{cases} \rho_n(w) & \text{if } b = B[i] \\ 0 & \text{if } b \neq B[i]. \end{cases}$$

If $|w| < 2^{d(n)}$, then $\rho_n(w)$ is computable in $O(|w|)$ time. If $|w| \geq 2^{d(n)}$, we can compute $\rho_n(w)$ by using $2^{n+1} - 1 = O(|w|^{n/d(n)}) = O(|w|^r)$ computations of $M$ and $f$ on strings with length at most $n = O(\log |w|)$. Therefore $\rho_n(w)$ is computable in $O(|w|^r(2^{c\log|w|} + (\log|w|)^k)) = O(|w|^{r+c})$ time for all $w \in \{0,1\}^*$.

Let $w_n = C \upharpoonright 2^{n^k+1} - 1$ be the characteristic string of $C_{\leq n^k}$. Then letting

$$m(n) = \left|\left\{j < |w_n| \,\big|\, (\forall i < 2^{n+1} - 1) f(s_i) \neq f(s_j)\right\}\right|,$$

we have

$$\rho_n(w_n) \geq \rho_n(\lambda) 2^{-2^{d(n)} - m(n)} = 2^{-2^{d(n)} - m(n) - n}.$$

By (6.1) and (6.2), we have

$$m(n) \leq 2^{n^k+1} - 1 - g_1(2^n, \alpha)$$

if $n$ is sufficiently large. In this case,

$$\log \rho_n(w_n) \geq g_1(2^n, \alpha) - 2^{d(n)} - 2^{n^k+1} - n. \tag{6.3}$$

The function $\rho : \{0,1\}^* \to [0,\infty)$ defined by

$$\rho(w) = \sum_{n=0}^\infty \rho_n(w)$$

for all $w$ is a measure by linearity. Notice that $\rho(w)$ can be approximated to a precision of $2^{-l}$ in $O(|w|^{r+c}l)$ time by adding the first $l+1$ terms of the sum.

Using (6.3), for all sufficiently large $n$, we have

$$\begin{aligned}
g_{-3}(|w_n|, s) + \log \rho_n(w_n) &= 2^{n^k+1} + 4 - g_3(2^{n^k+1} - 1, 1 - s) + \log \rho_n(w_n) \\
&\geq g_1(2^n, \alpha) - g_3(2^{n^k+1} - 1, 1 - s) - 2^{d(n)} - n.
\end{aligned}$$

By Lemma 6.1, $g_3(2^{n^k+1} - 1, 1 - s) = o(g_1(2^n, \alpha))$. Also, $2^{d(n)} = 2^{\lfloor n/r \rfloor}$ is little-$o$ of $g_1(2^n, \alpha) = 2^{\alpha n}$ because $\alpha > 1/r$. Using these facts, it follows that

$$\limsup_{n \to \infty} g_{-3}(n, s) + \log \rho_n(C \upharpoonright n) = \infty.$$

Appealing to Theorem 4.6, we establish $\dim_{\mathrm{p}}^{(i)}(\mathrm{P}_{\mathrm{m}}^{-1}(B)) \leq s$. As $s > 0$ is arbitrary, the $\Delta = \mathrm{p}$ part of the theorem holds. The argument is identical for $\Delta = \mathrm{pspace}$.

The proof for $\Delta \in \{\mathrm{p}_2, \mathrm{p}_2\mathrm{space}\}$ is very similar, so we only sketch the differences for $\Delta = \mathrm{p}_2$. Let $B \in \mathrm{EXP} - C_\alpha^{(2)}(2^n)$ and let $M$ be a Turing machine that decides $B$ in $O(2^{n^c})$ time. Assume $B \leq_{\mathrm{m}}^{\mathrm{P}} C$ via $f$. The measures $\rho_n$ and $\rho$ are defined in the same way, except we use a different function $d(n)$. For this, we let $r > 1/\alpha$ and define $d(n) = \lfloor n^\epsilon \rfloor$ where $\epsilon = 1/r$. Then, if $|w| \geq 2^{d(n)}$, as before we can compute $\rho_n(w)$ by using $2^{n+1} - 1$ computations of $M$ and $f$ on strings with length at most $n = O(\log |w|)$. Since $2^n = 2^{(\log 2^{n^\epsilon})^r} = O(2^{(\log |w|)^r})$, we can compute $\rho_n(w)$ in $O(2^{(\log |w|)^r} \cdot 2^{(\log |w|)^c}) = O(2^{(\log |w|)^{\max(r,c)}})$ time. Instead of (6.3), we arrive at $\log \rho_n(w_n) \geq g_2(2^n, \alpha) - 2^{d(n)} - 2^{n^k+1} - n$. The proof is completed in the same way using the fact that $2^{d(n)} = o(g_2(2^n, \alpha))$ because $\epsilon < \alpha$. $\square$

We are now ready to prove our main theorem.

**Theorem 6.3.**

1. *Let $\Delta \in \{\mathrm{p}, \mathrm{pspace}\}$. For every $A \in R(\Delta)$,*

$$\dim^{(1)}(\mathrm{P}_{\mathrm{m}}(A) \mid R(\Delta)) = 0$$

*or*

$$\dim^{(-3)}(\mathrm{P}_{\mathrm{m}}^{-1}(A) \mid R(\Delta)) = \dim_\Delta^{(-3)}(\mathrm{P}_{\mathrm{m}}^{-1}(A)) = 0.$$

2. *Let $\Delta \in \{\mathrm{p}_2, \mathrm{p}_2\mathrm{space}\}$. For every $A \in R(\Delta)$,*

$$\dim^{(2)}(\mathrm{P}_{\mathrm{m}}(A) \mid R(\Delta)) = 0$$

*or*

$$\dim^{(-3)}(\mathrm{P}_{\mathrm{m}}^{-1}(A) \mid R(\Delta)) = \dim_\Delta^{(-3)}(\mathrm{P}_{\mathrm{m}}^{-1}(A)) = 0.$$

*Proof.* Let $\Delta \in \{\mathrm{p}, \mathrm{pspace}\}$ and let $A \in R(\Delta)$. We consider two cases.

(I.) Suppose that

$$\mathrm{P}_{\mathrm{m}}(A) \cap R(\Delta) \subseteq \bigcap_{\alpha \in (0,1)} C_\alpha^{(1)}(2^n).$$

Then $\dim_\Delta^{(1)}(\mathrm{P}_{\mathrm{m}}(A) \cap R(\Delta)) \leq \dim_{\mathrm{p}}^{(1)}(C_\alpha^{(1)}(2^n)) \leq \alpha$ by Theorem 5.2 for all $\alpha \in (0,1)$, so $\dim^{(1)}(\mathrm{P}_{\mathrm{m}}(A) \mid R(\Delta)) = \dim_\Delta^{(1)}(\mathrm{P}_{\mathrm{m}}(A) \cap R(\Delta)) = 0$.

(II.) Otherwise, there is an $\alpha \in (0, 1)$ such that

$$\mathrm{P_m}(A) \cap R(\Delta) \not\subseteq C^{(1)}_\alpha(2^n).$$

Let $B \in \mathrm{P_m}(A) \cap R(\Delta) - C^{(1)}_\alpha(2^n)$. Then by Theorem 6.2, $\dim^{(-3)}_\Delta(\mathrm{P_m^{-1}}(B)) = 0$. Since $\mathrm{P_m^{-1}}(A) \subseteq \mathrm{P_m^{-1}}(B)$, we have $\dim^{(-3)}_\Delta(\mathrm{P_m^{-1}}(A)) = 0$.

Part 2 is proved in the same way. □

Theorem 6.3 implies that there is a small span theorem for $-3^{\mathrm{rd}}$-order scaled dimension, but it is stronger than the following.

**Corollary 6.4.** *For every $A \in R(\Delta)$,*

$$\dim^{(-3)}(\mathrm{P_m}(A) \mid R(\Delta)) = 0$$

*or*

$$\dim^{(-3)}(\mathrm{P_m^{-1}}(A) \mid R(\Delta)) = \dim^{(-3)}_\Delta(\mathrm{P_m^{-1}}(A)) = 0.$$

*Proof.* This follows immediately from Theorem 6.3 using Theorem 3.2. □

The small span theorem of Juedes and Lutz [10] is also a corollary.

**Corollary 6.5.** (Juedes and Lutz [10]) *Let $\Delta \in \{\mathrm{p}, \mathrm{p_2}\}$. For every $A \in R(\Delta)$,*

$$\mu(\mathrm{P_m}(A) \mid R(\Delta)) = 0$$

*or*

$$\mu(\mathrm{P_m^{-1}}(A) \mid R(\Delta)) = \mu_\Delta(\mathrm{P_m^{-1}}(A)) = 0.$$

*Proof.* This follows immediately from Theorem 6.3 and Lemma 3.1. □

We also have the following regarding the scaled dimensions of the hard languages for EXP and NP.

**Corollary 6.6.** *1. $\dim^{(-3)}_\mathrm{p}(\mathcal{H}^\mathrm{p}_\mathrm{m}(\mathrm{EXP})) = \dim^{(-3)}_{\mathrm{p_2}}(\mathcal{H}^\mathrm{p}_\mathrm{m}(\mathrm{EXP})) = 0$.*

*2. If $\dim^{(1)}(\mathrm{NP} \mid \mathrm{E}) > 0$, then $\dim^{(-3)}_\mathrm{p}(\mathcal{H}^\mathrm{p}_\mathrm{m}(\mathrm{NP})) = 0$.*

*3. If $\dim^{(2)}(\mathrm{NP} \mid \mathrm{EXP}) > 0$, then $\dim^{(-3)}_{\mathrm{p_2}}(\mathcal{H}^\mathrm{p}_\mathrm{m}(\mathrm{NP})) = 0$.*

*Proof.* Let $H \in \mathcal{C}^\mathrm{p}_\mathrm{m}(\mathrm{E})$. Then also $H \in \mathcal{C}^\mathrm{p}_\mathrm{m}(\mathrm{EXP})$, so $\mathrm{P_m^{-1}}(H) = \mathcal{H}^\mathrm{p}_\mathrm{m}(\mathrm{EXP})$. Since $\dim(\mathrm{P_m}(H) \mid \mathrm{E}) = \dim_\mathrm{p}(\mathrm{E}) = 1$, Theorem 6.3 tell us that $\dim_\mathrm{p}(\mathcal{H}^\mathrm{p}_\mathrm{m}(\mathrm{EXP})) = \dim_\mathrm{p}(\mathrm{P_m^{-1}}(H)) = 0$.

Parts 2 and 3 follow from Theorem 6.3 using any NP-complete language $A$. □

Juedes and Lutz [10] concluded from their small span theorem that *every $\leq^\mathrm{P}_\mathrm{m}$-degree has measure 0 in E and in EXP*. From Theorem 6.3 we similarly derive a stronger version of this fact: every $\leq^\mathrm{P}_\mathrm{m}$-degree actually has $-3^{\mathrm{rd}}$-order dimension 0.

**Corollary 6.7.** *For every $A \subseteq \{0,1\}^*$ and $\Delta \in \{\mathrm{p}, \mathrm{p_2}, \mathrm{pspace}, \mathrm{p_2space}\}$,*

$$\dim^{(-3)}(\deg^\mathrm{p}_\mathrm{m}(A) \mid R(\Delta)) = 0.$$

*Proof.* If $\deg_{\mathrm{m}}^{\mathrm{p}}(A)$ is disjoint from $R(\Delta)$, then $\dim^{(-3)}(\deg_{\mathrm{m}}^{\mathrm{p}}(A) \mid R(\Delta)) = \dim_{\mathrm{p}}^{(-3)}(\emptyset) = 0$, so assume that there is some $B \in \deg_{\mathrm{m}}^{\mathrm{p}}(A) \cap R(\Delta)$. Because $\deg_{\mathrm{m}}^{\mathrm{p}}(A) = \deg_{\mathrm{m}}^{\mathrm{p}}(B) = \mathrm{P}_{\mathrm{m}}(B) \cap \mathrm{P}_{\mathrm{m}}^{-1}(B)$, we have

$$\dim^{(-3)}(\deg_{\mathrm{m}}^{\mathrm{p}}(A) \mid R(\Delta)) \leq \dim^{(-3)}(\mathrm{P}_{\mathrm{m}}(B) \mid R(\Delta))$$

and

$$\dim^{(-3)}(\deg_{\mathrm{m}}^{\mathrm{p}}(A) \mid R(\Delta)) \leq \dim^{(-3)}(\mathrm{P}_{\mathrm{m}}^{-1}(B) \mid R(\Delta)).$$

By Corollary 6.4, we have either $\dim^{(-3)}(\mathrm{P}_{\mathrm{m}}(B) \mid R(\Delta)) = 0$ or $\dim^{(-3)}(\mathrm{P}_{\mathrm{m}}^{-1}(B) \mid R(\Delta)) = 0$. Therefore $\dim^{(-3)}(\deg_{\mathrm{m}}^{\mathrm{p}}(A) \mid R(\Delta)) = 0$. □

The $\leq_{\mathrm{m}}^{\mathrm{P}}$-complete languages for any complexity class has $-3^{\mathrm{rd}}$-order dimension in every $R(\Delta)$.

**Corollary 6.8.** *For any class $\mathcal{D}$ of languages, $\dim^{(-3)}(\mathcal{C}_{\mathrm{m}}^{\mathrm{p}}(\mathcal{D}) \mid R(\Delta)) = 0$.*

*Proof.* If $\mathcal{C}_{\mathrm{m}}^{\mathrm{p}}(\mathcal{D}) = \emptyset$, this is trivial. Assume $\mathcal{C}_{\mathrm{m}}^{\mathrm{p}}(\mathcal{D}) \neq \emptyset$ and let $A \in \mathcal{C}_{\mathrm{m}}^{\mathrm{p}}(\mathcal{D})$. Then $\mathcal{C}_{\mathrm{m}}^{\mathrm{p}}(\mathcal{D}) \subseteq \deg_{\mathrm{m}}^{\mathrm{p}}(A)$, so this follows from Corollary 6.7. □

## 7 Lower Spans vs. Degrees in Orders -2 Through 2

We now present some results that stand in contrast to the small span theorem of the previous section. We begin by showing that lower spans and degrees have the same scaled dimension in orders $i$ with $|i| \leq 2$.

**Theorem 7.1.** *For any $A \in \mathrm{E}$, $-2 \leq i \leq 2$, and $\Delta \in \{\mathrm{p}, \mathrm{p}_2, \mathrm{pspace}, \mathrm{p}_2\mathrm{space}\}$,*

$$\dim^{(i)}(\deg_{\mathrm{m}}^{\mathrm{p}}(A) \mid R(\Delta)) = \dim^{(i)}(\mathrm{P}_{\mathrm{m}}(A) \mid R(\Delta))$$

*and*

$$\dim_{\Delta}^{(i)}(\deg_{\mathrm{m}}^{\mathrm{p}}(A)) = \dim_{\Delta}^{(i)}(\mathrm{P}_{\mathrm{m}}(A)).$$

*Proof.* We write the proof for dimension in $R(\mathrm{p}) = \mathrm{E}$; the rest of theorem is proved in the same manner.

Let $A \in \mathrm{E}$ be decidable in $O(2^{cn})$ time. By monotonicity, $\dim^{(i)}(\deg_{\mathrm{m}}^{\mathrm{p}}(A) \mid \mathrm{E}) \leq \dim^{(i)}(\mathrm{P}_{\mathrm{m}}(A) \mid \mathrm{E})$. For the other inequality, let $t > s > \dim^{(i)}(\deg_{\mathrm{m}}^{\mathrm{p}}(A) \mid \mathrm{E})$. By Lemmas 4.2 and 4.3, for some $l \in \mathbb{N}$ there is an exactly $n^l$-time computable measure $\rho$ satisfying

$$\limsup_{m \to \infty} g_i(m, s) + \log \rho(C \upharpoonright m) = \infty \tag{7.1}$$

for all $C \in \deg_{\mathrm{m}}^{\mathrm{p}}(A) \cap \mathrm{E}$.

Letting $k \geq 1$ be a natural number to be specified later, we define a padding function $f : \{0,1\}^* \to \{0,1\}^*$ by

$$f(x) = 0^{|x|^k - |x|} x$$

for all $x$. Let $R = f(\{0,1\}^*)$ be the range of $f$.

Let $B \in \mathrm{P}_{\mathrm{m}}(A)$. We define another language $B'$ as

$$B' = (B - R) \cup f(A).$$

Then $B' \in \deg_m^p(A)$. Intuitively, $B'$ is a language that is very similar to $B$ but has $A$ encoded sparsely in it. Define a function $\tau : \{0,1\}^* \to \{0,1\}^*$ inductively by $\tau(\lambda) = \lambda$ and

$$\tau(wb) = \begin{cases} \tau(w)b & \text{if } s_{|w|} \notin R \\ \tau(w)1 & \text{if } s_{|w|} \in R \cap B' \\ \tau(w)0 & \text{if } s_{|w|} \in R - B' \end{cases}$$

for all $w \in \{0,1\}^*$ and $b \in \{0,1\}$. Notice that

$$\tau(B \restriction n) = B' \restriction n$$

for all $n$.

Define a measure $\gamma$ as follows. For all $w \in \{0,1\}^*$ and $b \in \{0,1\}$,

$$\gamma(wb) = \begin{cases} \frac{\gamma(w)}{2} & \text{if } s_{|w|} \in R \\ \frac{\rho(\tau(w)b)}{\rho(\tau(w))} \gamma(w) & \text{if } s_{|w|} \notin R. \end{cases}$$

Intuitively, $\gamma$ is designed to have performance on $B$ that is similar to $\rho$'s performance on $B'$. This is done by mimicking the conditional probabilities of $\rho$ for strings that are not in $R$. Note that $\gamma(w)$ can be exactly computed in $O(|w| \cdot (|w|^l + 2^{(\log|w|)^c})) = O(|w|^{\max(l,c)})$ time.

Let $n \in N$ and let $2^{(n-1)^k+1} \leq m \leq 2^{n^k+1} - 1$. Then

$$\begin{aligned}
\log \gamma(B \restriction m) &= \sum_{1 \leq i \leq m} \log \frac{\gamma(B \restriction i)}{\gamma(B \restriction i-1)} \\
&= \sum_{\substack{1 \leq i \leq m \\ s_i \notin R}} \log \frac{\rho(\tau(B \restriction i-1)B[i])}{\rho(\tau(B \restriction i-1))} + \sum_{\substack{1 \leq i \leq m \\ s_i \in R}} \log \frac{1}{2} \\
&= \sum_{\substack{1 \leq i \leq m \\ s_i \notin R}} \log \frac{\rho(B' \restriction i)}{\rho(B' \restriction i-1)} - |\{1 \leq i \leq m \mid s_i \in R\}| \\
&\geq \sum_{1 \leq i \leq m} \log \frac{\rho(B' \restriction i)}{\rho(B' \restriction i-1)} - |\{1 \leq i \leq 2^{n^k+1} - 1 \mid s_i \in R\}| \\
&= \log \rho(B' \restriction m) - \sum_{i=0}^{n} 2^n \\
&= \log \rho(B' \restriction m) - 2^{n+1} + 1.
\end{aligned}$$

Now assume that $g_i(m, s) + \log \rho(B' \restriction m) \geq 1$. Then we have $g_i(m, t) + \log \gamma(B \restriction m) \geq 1$ if

$$2^{n+1} + g_i(m, s) < g_i(m, t). \tag{7.2}$$

To establish

$$\limsup_{n \to \infty} g_i(m, t) + \log \gamma(B \restriction m) \geq 1, \tag{7.3}$$

it now suffices to show that (7.2) holds for all sufficiently large $m$. For each $-2 \leq i \leq 2$, we now give an appropriate choice of $k$ that yields this.

- **i = 2**: Let $k > 1/t$. Then $g_2(m,t) \geq g_2(2^{(n-1)^k}, t) = 2^{(n-1)^{kt}}$, so $2^{n+1} = o(g_2(m,t))$ because $kt > 1$. Also, $g_2(m,s) = o(g_2(m,t))$ since $s < t$, so (7.2) holds when $m$ is sufficiently large.

- **i = 1**: Let $k = 2$. Then $g_1(m,t) \geq g_1(2^{(n-1)^2}, t) = 2^{t(n-1)^2}$, so $2^{n+1} = o(g_2(m,t))$. Also, $g_1(m,s) = o(g_1(m,t))$, so (7.2) holds for sufficiently large $m$.

- **i = 0**: Let $k = 2$. Then $g_0(m,t) \geq g_0(2^{(n-1)^2}, t) = t2^{(n-1)^2}$, so $2^{n+1} = o(g_0(m,t))$. Also, $g_0(m,s) = o(g_0(m,t))$, so (7.2) holds for sufficiently large $m$.

- **i = −1**: We have $g_{-1}(m,t) = m + 1 - g_1(m, 1-t)$, so (7.2) is true if $2^{n+1} + g_1(m, 1-t) < g_1(m, 1-s)$. Taking $k = 2$, this follows from the argument for $i = 1$ above since $1 - s > 1 - t$.

- **i = −2**: Just as in the $i = -1$ case, (7.2) is true if $2^{n+1} + g_2(m, 1-t) < g_2(m, 1-s)$. Taking $k > 1/(1-s)$, this follows from the argument for $i = 2$ above since $1 - s > 1 - t$.

For each $B \in \mathrm{P_m}(A)$, we have given a $O(n^{\max(l,c)})$-time computable measure $\gamma$ such that (7.3) holds. By Theorem 4.6, $\dim^{(i)}(\mathrm{P_m}(A) \mid \mathrm{E}) \leq t$. As $t > \dim^{(i)}(\deg_{\mathrm{m}}^{\mathrm{P}}(A) \mid \mathrm{E})$ is arbitrary, this establishes $\dim^{(i)}(\mathrm{P_m}(A) \mid \mathrm{E}) \leq \dim^{(i)}(\deg_{\mathrm{m}}^{\mathrm{P}}(A) \mid \mathrm{E})$. □

Theorem 7.1 for (unscaled) dimension was proved in [2] for $\Delta = \mathrm{p}$.

**Corollary 7.2.** (Ambos-Spies, Merkle, Reimann, and Stephan [2]) *For any $A \in \mathrm{E}$,*

$$\dim(\deg_{\mathrm{m}}^{\mathrm{P}}(A) \mid \mathrm{E}) = \dim(\mathrm{P_m}(A) \mid \mathrm{E})$$

*and*

$$\dim_{\mathrm{p}}(\deg_{\mathrm{m}}^{\mathrm{P}}(A)) = \dim_{\mathrm{p}}(\mathrm{P_m}(A)).$$

Theorem 7.1 implies that Theorem 6.3 cannot be improved in one respect. For any $i, j \in \mathbb{Z}$, let $\mathrm{SST}[i,j]$ be the assertion that for every $A \in \mathrm{E}$, either

$$\dim^{(i)}(\mathrm{P_m}(A) \mid \mathrm{E}) = 0$$

or

$$\dim^{(j)}(\mathrm{P_m}^{-1}(A) \mid \mathrm{E}) = 0.$$

Let $H \in \mathcal{C}_{\mathrm{m}}^{\mathrm{P}}(\mathrm{E})$. Then

$$\dim^{(-2)}(\mathrm{P_m}(H) \mid \mathrm{E}) = \dim^{(-2)}(\mathrm{E} \mid \mathrm{E}) = 1,$$

so $\dim^{(-2)}(\deg_{\mathrm{m}}^{\mathrm{P}}(H) \mid \mathrm{E}) = 1$ by Theorem 7.1, which in turn implies

$$\dim^{(-2)}(\mathrm{P_m}^{-1}(H) \mid \mathrm{E}) = 1.$$

Therefore, $\mathrm{SST}[i,j]$ is true only if $i \leq -3$ or $j \leq -3$. Theorem 6.3 asserts $\mathrm{SST}[1, -3]$, so the $-3$ in it cannot be improved to $-2$.

We have the following corollary regarding the classes of complete sets for E, EXP, and NP.

**Corollary 7.3.** *Let $-2 \leq i \leq 2$.*

1. $\dim^{(i)}(\mathcal{C}_{\mathrm{m}}^{\mathrm{P}}(\mathrm{E}) \mid \mathrm{E}) = \dim^{(i)}(\mathcal{C}_{\mathrm{m}}^{\mathrm{P}}(\mathrm{EXP}) \mid \mathrm{EXP}) = 1$.

2. $\dim^{(i)}(\mathrm{NP} \mid \mathrm{E}) = \dim^{(i)}(\mathcal{C}_{\mathrm{m}}^{\mathrm{P}}(\mathrm{NP}) \mid \mathrm{E})$.

3. $\dim^{(i)}(\mathrm{NP} \mid \mathrm{EXP}) = \dim^{(i)}(\mathcal{C}_\mathrm{m}^\mathrm{p}(\mathrm{NP}) \mid \mathrm{EXP})$.

*Proof.* Let $H \in \mathcal{C}_\mathrm{m}^\mathrm{p}(\mathrm{E})$. Then $\mathcal{C}_\mathrm{m}^\mathrm{p}(\mathrm{E}) = \deg_\mathrm{m}^\mathrm{p}(H) \cap \mathrm{E}$, so $\dim^{(i)}(\mathcal{C}_\mathrm{m}^\mathrm{p}(\mathrm{E}) \mid \mathrm{E}) = \dim^{(i)}(\deg_\mathrm{m}^\mathrm{p}(H) \mid \mathrm{E}) = \dim^{(i)}(\mathrm{P}_\mathrm{m}(H) \mid \mathrm{E}) = \dim_\mathrm{p}^{(i)}(\mathrm{E}) = 1$ by Theorem 7.1. The other statements follow similarly. □

We can now observe a difference between the $-3^\mathrm{rd}$- and $-2^\mathrm{nd}$-order scaled dimensions regarding complete degrees. Corollaries 6.8 and 7.3 together with Theorem 3.2 tell us that for $\mathcal{D} \in \{\mathrm{E}, \mathrm{EXP}\}$,

$$\dim^{(i)}(\mathcal{C}_\mathrm{m}^\mathrm{p}(\mathcal{D}) \mid \mathcal{D}) = \begin{cases} 0 & \text{if } i \leq -3 \\ 1 & \text{if } i \geq -2 \end{cases}$$

and

$$\dim^{(i)}(\mathcal{C}_\mathrm{m}^\mathrm{p}(\mathrm{NP}) \mid \mathcal{D}) = \begin{cases} 0 & \text{if } i \leq -3 \\ \dim^{(i)}(\mathrm{NP} \mid \mathcal{D}) & \text{if } i \geq -2. \end{cases}$$

In Section 9 we will discuss the scaled dimension of $\mathcal{C}_\mathrm{m}^\mathrm{p}(\mathrm{E})$ within ESPACE. The following extension of Theorem 7.1 will be useful.

**Theorem 7.4.** *For all $-2 \leq i \leq 2$,*

$$\dim^{(i)}(\mathcal{C}_\mathrm{m}^\mathrm{p}(\mathrm{E}) \mid \mathrm{ESPACE}) = \dim^{(i)}(\mathrm{E} \mid \mathrm{ESPACE}).$$

*Proof.* We use the construction from the proof of Theorem 7.1. Let $t > s > \dim^{(i)}(\mathcal{C}_\mathrm{m}^\mathrm{p}(\mathrm{E}) \mid \mathrm{ESPACE})$ and take an exactly $n^l$-space computable measure $\rho$ satisfying (7.1) for all $C \in \mathcal{C}_\mathrm{m}^\mathrm{p}(\mathrm{E})$. Fix an $A \in \mathcal{C}_\mathrm{m}^\mathrm{p}(\mathrm{E})$. For any $B \in \mathrm{E}$, the set $B'$ constructed from $A$ and $B$ is in $\mathcal{C}_\mathrm{m}^\mathrm{p}(\mathrm{E})$. The arguments then show $\dim^{(i)}(\mathrm{E} \mid \mathrm{ESPACE}) \leq t$. □

# 8 $\leq_{1-\mathrm{tt}}^\mathrm{P}$-Lower Spans vs. $\leq_\mathrm{m}^\mathrm{P}$-Lower Spans

Theorem 7.1 is also true for most other polynomial-time reducibilities. (This fact was mentioned in [2] for Corollary 7.2 when it was proved.) To replace $\leq_\mathrm{m}^\mathrm{P}$ by $\leq_r^\mathrm{P}$ in the theorem, we only need to have $B' \in \deg_r^\mathrm{p}(A)$ for the set $B'$ that was constructed in the proof from $B \in \mathrm{P}_r(A)$. In particular, Theorem 7.1 is true for the $\leq_{1-\mathrm{tt}}^\mathrm{P}$ reducibility. In this section we show that this holds because of another reason: the scaled dimensions of $\leq_{1-\mathrm{tt}}^\mathrm{P}$-lower spans and $\leq_\mathrm{m}^\mathrm{P}$-lower spans are always the same.

The following proposition was used to show that a set is weakly $\leq_\mathrm{m}^\mathrm{P}$-complete for exponential time if and only if it is $\leq_{1-\mathrm{tt}}^\mathrm{P}$-complete.

**Proposition 8.1.** *(Ambos-Spies, Mayordomo, and Zheng [1]) Let $A \leq_{1-\mathrm{tt}}^\mathrm{P} B$. Then there is a language $C \in \mathrm{P}$ such that*

$$\hat{A} = (A \cap C) \cup (A^c \cap C^c) \leq_\mathrm{m}^\mathrm{P} B.$$

The idea of the following lemma also comes from [1].

**Lemma 8.2.** *Let $i \in \mathbb{Z}$. Let $\mathcal{C}, \hat{\mathcal{C}}$ be classes of languages such that for any $A \in \mathcal{C}$, there is some $C \in R(\Delta)$ such that $\hat{A} = (A \cap C) \cup (A^c \cap C^c) \in \hat{\mathcal{C}}$. Then $\dim_\Delta^{(i)}(\mathcal{C}) \leq \dim_\Delta^{(i)}(\hat{\mathcal{C}})$.*

*Proof.* We prove this for $\Delta = \mathrm{p}$. The other cases are proved by identical arguments.

Let $s > \dim_{\mathrm{p}}^{(i)}(\hat{\mathcal{C}})$ be rational and obtain $\rho$ computable in $O(n^r)$ time from Lemma 4.2 such that
$$\limsup_{n \to \infty} g_i(n, s) + \log \rho(\hat{A} \upharpoonright n) = \infty \qquad (8.1)$$
for all $\hat{A} \in \hat{\mathcal{C}}$.

Let $A \in \mathcal{C}$ and let $C \in \mathrm{DTIME}(n^k)$ such that $\hat{A} = (A \cap C) \cup (A^c \cap C^c) \in \hat{\mathcal{C}}$. Define a function $\tau : \{0,1\}^* \to \{0,1\}^*$ by
$$\tau(w)[j] = \begin{cases} w[j] & \text{if } s_j \in C \\ 1 - w[j] & \text{if } s_j \notin C \end{cases}$$
for each $0 \leq j < |w|$. Define another measure $\rho'$ by
$$\rho'(w) = \rho(\tau(w)).$$

Then for all $n$,
$$\rho'(A \upharpoonright n) = \rho(\tau(A \upharpoonright n)) = \rho(\hat{A} \upharpoonright n).$$

Therefore
$$\limsup_{n \to \infty} g_i(n, s) + \log \rho'(A \upharpoonright n) = \infty$$
because of (8.1). As $\rho'$ is computable in time $O(|w| \cdot (\log |w|)^k + |w|^r) = O(|w|^2 + |w|^r)$, it follows by Theorem 4.6 that $\dim_{\mathrm{p}}^{(i)}(\mathcal{C}) \leq s$. $\square$

We now show that the scaled dimension of a $\leq_{\mathrm{m}}^{\mathrm{P}}$-lower span is always equal to the scaled dimension of the $\leq_{1-\mathrm{tt}}^{\mathrm{P}}$-lower span.

**Theorem 8.3.** *Let $B \subseteq \{0,1\}^*$ and let $i \in \mathbb{Z}$. Then*
$$\dim_{\Delta}^{(i)}(\mathrm{P}_{\mathrm{m}}(B)) = \dim_{\Delta}^{(i)}(\mathrm{P}_{1-\mathrm{tt}}(B))$$
*and*
$$\dim^{(i)}(\mathrm{P}_{\mathrm{m}}(B) \mid R(\Delta)) = \dim^{(i)}(\mathrm{P}_{1-\mathrm{tt}}(B) \mid R(\Delta)).$$

*Proof.* By Proposition 8.1, for each $A \in \mathrm{P}_{1-\mathrm{tt}}(B)$ there is a language $C \in \mathrm{P}$ such that $\hat{A} = (A \cap C) \cup (A^c \cap C^c) \in \mathrm{P}_{\mathrm{m}}(B)$. Let $\hat{\mathcal{C}}$ be the set of all such $\hat{A}$ as $A$ ranges over $\mathrm{P}_{1-\mathrm{tt}}(B)$. Then by Lemma 8.2,
$$\dim_{\Delta}^{(i)}(\mathrm{P}_{1-\mathrm{tt}}(B)) \leq \dim_{\Delta}^{(i)}(\hat{\mathcal{C}}).$$

As $\hat{\mathcal{C}} \subseteq \mathrm{P}_{\mathrm{m}}(B) \subseteq \mathrm{P}_{1-\mathrm{tt}}(B)$, we also have
$$\dim_{\Delta}^{(i)}(\hat{\mathcal{C}}) \leq \dim_{\Delta}^{(i)}(\mathrm{P}_{\mathrm{m}}(B)) \leq \dim_{\Delta}^{(i)}(\mathrm{P}_{1-\mathrm{tt}}(B)),$$
so the first equality holds. The proof for dimension in $R(\Delta)$ is analogous. $\square$

We can now give a stronger version of Theorem 7.1.

**Corollary 8.4.** *For any $A \in R(\Delta)$ and $-2 \leq i \leq 2$,*

$$\begin{array}{ccc}
\dim^{(i)}(\mathrm{P_m}(A) \mid R(\Delta)) & = & \dim^{(i)}(\deg^{\mathrm{p}}_{\mathrm{m}}(A) \mid R(\Delta)) \\
\| & & \| \\
\dim^{(i)}(\mathrm{P_{1-tt}}(A) \mid R(\Delta)) & = & \dim^{(i)}(\deg^{p}_{1-\mathrm{tt}}(A) \mid R(\Delta)),
\end{array}$$

*and similarly with $\dim^{(i)}(\cdot \mid R(\Delta))$ replaced by $\dim^{(i)}_\Delta(\cdot)$.*

*Proof.* From Theorems 7.1 and 8.3 we have

$$\dim^{(i)}(\deg^{\mathrm{p}}_{\mathrm{m}}(A) \mid R(\Delta)) = \dim^{(i)}(\mathrm{P_m}(A) \mid R(\Delta)) = \dim^{(i)}(\mathrm{P_{1-tt}}(A) \mid R(\Delta)).$$

By monotonicity, we have

$$\dim^{(i)}(\deg^{\mathrm{p}}_{\mathrm{m}}(A) \mid R(\Delta)) \leq \dim^{(i)}(\deg^{\mathrm{p}}_{1-\mathrm{tt}}(A) \mid R(\Delta)) \leq \dim^{(i)}(\mathrm{P_{1-tt}}(A) \mid R(\Delta)),$$

so the equalities displayed in the statement of the corollary are true. The proof for $\dim^{(i)}_\Delta(\cdot)$ is analogous. □

Theorem 8.3 also yields a strengthening of Theorem 6.3: the $\mathrm{P_m}(A)$ in it can be replaced by $\mathrm{P_{1-tt}}(A)$. In fact, it is also possible to also replace the $\mathrm{P}^{-1}_{\mathrm{m}}(A)$ in Theorem 6.3 by $\mathrm{P}^{-1}_{1-\mathrm{tt}}(A)$ by extending Theorems 5.2 and 6.2 to deal with $\leq^{\mathrm{P}}_{1-\mathrm{tt}}$-reductions. We omit the details.

## 9 The Scaled Dimension of $\mathcal{C}^{\mathrm{p}}_{\mathrm{m}}(\mathrm{E})$ in ESPACE

Lutz [15] proved a small span theorem for nonuniform Turing reductions in ESPACE. This implies that $\mathcal{C}^{\mathrm{p}}_{\mathrm{m}}(\mathrm{E})$ has measure 0 in ESPACE. In Corollary 6.8 we saw that $\mathcal{C}^{\mathrm{p}}_{\mathrm{m}}(\mathrm{E})$ actually has $-3^{\mathrm{rd}}$-order scaled dimension 0 in ESPACE. In this section we show that determining the $-2^{\mathrm{nd}}$- or $-1^{\mathrm{st}}$-order scaled dimension of $\mathcal{C}^{\mathrm{p}}_{\mathrm{m}}(\mathrm{E})$ in ESPACE would yield a proof of P = BPP or P $\neq$ PSPACE.

The P = BPP hypothesis was related to the measure of E in ESPACE by Lutz [13].

**Theorem 9.1.** (Lutz [13]) *If $\mu(\mathrm{E} \mid \mathrm{ESPACE}) \neq 0$, then P = BPP.*

We will extend this result to scaled dimension. We now recall the tools Lutz used to proved it.

Nisan and Wigderson [19] showed that BPP can be derandomized if there is a decision problem in E that requires exponential-size circuits to approximately solve. The *hardness* of a decision problem at a given length is the minimum size of a circuit that can approximately solve it. The details of the definition of this hardness are not needed in this paper; we only need to recall existing results regarding classes of languages with exponential hardness.

**Definition.** *Let $H_\alpha$ be the class of all languages that have hardness at least $2^{\alpha n}$ almost everywhere.*

The aforementioned derandomization of BPP can be stated as follows.

**Theorem 9.2.** (Nisan and Wigderson [19]) *If there is an $\mathrm{E} \cap H_\alpha \neq \emptyset$ for some $\alpha > 0$, then P = BPP.*

We will also need space-bounded Kolmogorov complexity.

**Definition.** *Given a machine $M$, a space bound $s : \mathbb{N} \to \mathbb{N}$, a language $L \subseteq \{0,1\}^*$, and a natural number $n$, the $s$-space-bounded Kolmogorov complexity of $L_{=n}$ relative to $M$ is*

$$\mathrm{KS}^t_M(L_{=n}) = \min\left\{|\pi| \,\Big|\, M(\pi, n) = \chi_{L_{=n}} \text{ in } \leq t(2^n) \text{ space}\right\},$$

*i.e., the length of the shortest program $\pi$ such that $M$, on input $(\pi, n)$, outputs the characteristic string of $L_{=n}$ and halts without using more than $t(2^n)$ workspace.*

Well-known simulation techniques show that there exists a machine $U$ which is *optimal* in the sense that for each machine $M$ there is a constant $c$ such that for all $t$, $L$ and $n$ we have

$$\mathrm{KS}^{ct+c}_U(L_{=n}) \leq \mathrm{KS}^t_M(L_{=n}) + c.$$

As usual, we fix such a universal machine and omit it from the notation.

**Definition.** *For each space bound $s : \mathbb{N} \to \mathbb{N}$ and function $f : \mathbb{N} \to \mathbb{N}$ define the complexity class*

$$\mathrm{KS}^t(f) = \{L \subseteq \{0,1\}^* \mid (\forall^\infty n)\mathrm{KS}^t(L_{=n}) < f(n)\}.$$

Lutz showed that $H_\alpha$ has measure 1 in ESPACE (i.e., that $H_\alpha^c$ has measure 0 in ESPACE) if $\alpha < 1/3$ by showing that languages not in $H_\alpha$ have low space-bounded Kolmogorov complexity.

**Lemma 9.3.** (Lutz [13]) *There exist a polynomial $q$ and a constant $c$ such that for all $0 < \alpha < \beta < 1$,*

$$H_\alpha^c \subseteq \mathrm{KS}^q_{\mathrm{i.o.}}(2^n - c2^{(1-2\alpha)n} + 2^{\beta n}).$$

The class on the right in Lemma 9.3 has measure 0 in ESPACE [14]. The scaled dimensions of similar space-bounded Kolmogorov complexity classes were studied in [9].

**Theorem 9.4.** (Hitchcock, Lutz, and Mayordomo [9]) *For any $i \leq -1$, polynomial $q(n) = \Omega(n^2)$, and $\alpha \in [0,1]$,*

$$\dim^{(i)}(\mathrm{KS}^q_{\mathrm{i.o.}}(g_i(2^n, \alpha)) \mid \mathrm{ESPACE}) = \alpha.$$

Lemma 9.3 and Theorem 9.4 provide an easy upper bound on the $-1^{\mathrm{st}}$-order scaled dimension of $H_\alpha^c$ in ESPACE.

**Corollary 9.5.** *If $0 < \alpha < 1/3$, then*

$$\dim^{(-1)}_{\mathrm{pspace}}(H_\alpha^c) \leq 2\alpha.$$

*Proof.* Let $\epsilon > 0$ and $\beta \in (\alpha, 1 - 2\alpha)$. Then for all sufficiently large $n$,

$$2^n - c2^{(1-2\alpha)n} + 2^{\beta n} < 2^n + 1 - 2^{(1-2\alpha-\epsilon)n}$$
$$= g_1(2^n, 2\alpha + \epsilon),$$

so Lemma 9.3 implies $H_\alpha^c \subseteq \mathrm{KS}^q_{\mathrm{i.o.}}(g_1(2^n, 2\alpha + \epsilon))$. Therefore $\dim^{(-1)}(H_\alpha^c \mid \mathrm{ESPACE}) \leq 2\alpha + \epsilon$ by Theorem 9.4. □

We can now state a stronger version of Theorem 9.1. The hypothesis has been weakened, but the conclusion remains the same.

**Theorem 9.6.** *If $\dim^{(-1)}(\mathrm{E} \mid \mathrm{ESPACE}) > 0$, then $\mathrm{P} = \mathrm{BPP}$.*

*Proof.* Assume the hypothesis and let $s = \min\{1/2, \dim^{(-1)}(\text{E} \mid \text{ESPACE})\}$. Then by Corollary 9.5, $\text{E} \not\subseteq H^c_{s/2}$, i.e., $\text{E} \cap H_{s/2} \neq \emptyset$. Therefore $\text{P} = \text{BPP}$ by Theorem 9.2. □

We now relate the scaled dimension of $\mathcal{C}^\text{p}_\text{m}(\text{E})$ to the $\text{P} \stackrel{?}{=} \text{PSPACE}$ and $\text{P} \stackrel{?}{=} \text{BPP}$ problems.

**Theorem 9.7.** *For $i \in \{-2, -1\}$,*
$$\dim^{(i)}(\mathcal{C}^\text{p}_\text{m}(\text{E}) \mid \text{ESPACE}) < 1 \Rightarrow \text{P} \neq \text{PSPACE}$$
*and*
$$\dim^{(i)}(\mathcal{C}^\text{p}_\text{m}(\text{E}) \mid \text{ESPACE}) > 0 \Rightarrow \text{P} = \text{BPP}.$$

*Proof.* From Theorem 7.4 we know that $\dim^{(i)}(\mathcal{C}^\text{p}_\text{m}(\text{E}) \mid \text{ESPACE}) = \dim^{(i)}(\text{E} \mid \text{ESPACE})$. Also, $\dim^{(i)}(\text{E} \mid \text{ESPACE}) < 1$ implies $\text{E} \neq \text{ESPACE}$ which is equivalent to $\text{P} \neq \text{PSPACE}$ [5]. This proves the first implication. The second one follows from Theorem 9.6 since $\dim^{(i)}(\mathcal{C}^\text{p}_\text{m}(\text{E}) \mid \text{ESPACE}) > 0$ implies $\dim^{(-1)}(\text{E} \mid \text{ESPACE}) > 0$. □

In other words, establishing any nontrivial upper or lower bound on $\dim^{(-1)}(\mathcal{C}^\text{p}_\text{m}(\text{E}) \mid \text{ESPACE})$ or $\dim^{(-2)}(\mathcal{C}^\text{p}_\text{m}(\text{E}) \mid \text{ESPACE})$ would derandomize BPP or separate P from PSPACE. This is in contrast to the unconditional facts from Corollaries 6.7 and 7.3 that
$$\dim^{(-3)}(\mathcal{C}^\text{p}_\text{m}(\text{E}) \mid \text{ESPACE}) = 0$$
and
$$\dim^{(-2)}(\mathcal{C}^\text{p}_\text{m}(\text{E}) \mid \text{E}) = \dim^{(-1)}(\mathcal{C}^\text{p}_\text{m}(\text{E}) \mid \text{E}) = 1.$$

## 10 Conclusion

Our main results, Theorems 6.3 and 7.1, use resource-bounded scaled dimension to strengthen from both ends the contrasting theorems of Juedes and Lutz [10] and Ambos-Spies, Merkle, Reimann, and Stephan [2] regarding spans under polynomial-time reductions.

1. The *small span theorem* for $\leq^\text{p}_\text{m}$-reductions [10] was strengthened from measure to $-3^\text{rd}$-order scaled dimension. (In fact, Theorem 6.3 is even stronger than this.)

2. The result that lower spans and degrees have the same dimension [2] was extended to all orders $-2 \leq i \leq 2$ of scaled dimension. This implies that there is no small span theorem in $-2^\text{nd}$-order scaled dimension.

These results suggest that contrast between the $-2^\text{nd}$- and $-3^\text{rd}$-orders of resource-bounded scaled dimension will be useful for studying complexity classes involving polynomial-time reductions. For example, regarding the many-one complete degree of NP, Corollaries 6.7 and 7.3 say that
$$\dim^{(-3)}(\mathcal{C}^\text{p}_\text{m}(\text{NP}) \mid \text{E}) = 0$$
and
$$\dim^{(-2)}(\mathcal{C}^\text{p}_\text{m}(\text{NP}) \mid \text{E}) = \dim^{(-2)}(\text{NP} \mid \text{E}).$$

Scaled dimension therefore provides two different types of dimension for studying NP. The NP-complete degree provides all the dimension of NP in order -2, but in order -3 the NP-complete degree unconditionally has dimension 0.